# Noise properties in the ideal Kirchhoff-Law-Johnson-Noise secure communication system

Zoltan Gingl and Robert Mingesz

Department of Technical Informatics, University of Szeged, Hungary

## *Abstract*

In this paper we determine the noise properties needed for unconditional security for the ideal Kirchhoff-Law-Johnson-Noise (KLJN) secure key distribution system using simple statistical analysis. It has already been shown using physical laws that resistors and Johnson-like noise sources provide unconditional security. However real implementations use artificial noise generators, therefore it is a question if other kind of noise sources and resistor values could be used as well. We answer this question and in the same time we provide a theoretical basis to analyze real systems as well.

## *Introduction*

Communication security is getting more and more important in many different applications including electronic banking, protecting personal data, securing intellectual property of companies, transmission of medical data and many more. The Kirchhoff-Law-Johnson-Noise (KLJN) protocol was introduced as a low cost unconditionally secure key exchange protocol using only passive components: four resistors, two switches and interconnecting wires [1]. The protocol is based only on the laws of classical physics and has been introduced as an inexpensive alternative to quantum communicators. The first real implementation has been shown a few years after its discovery [2, 3] and it has inspired the development of another secret key exchanged method [4]. There are many potential applications including securing computers, algorithms and hardware (memories, processors, keyboards, mass storage media) [5], key distribution over the Smart Grid [6], ethernet cables [7], uncloneable hardware keys [8]. Several attack methods has been discussed [9-14], however the ideal KLJN system is found to be secure. Debates are still going on [15,16] and recent papers discuss practical considerations for the applications [17,18].

The KLJN key exchange protocol is rather simple. During the communication a secret key is generated and shared between the two communicating parties, Alice and Bob. The system consists of two communicators and a transmission wire, see Fig. 1.

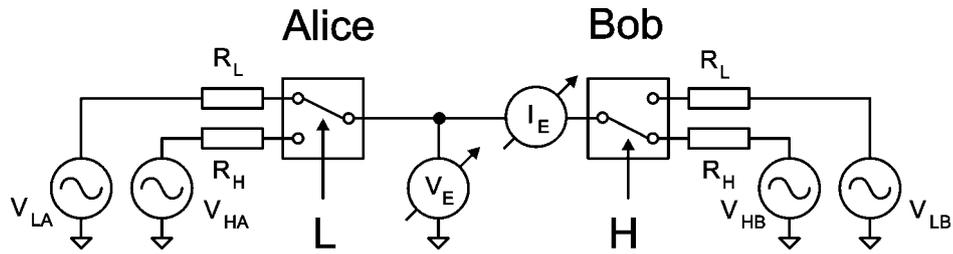

Figure 1. The KLJN secure communication system.

Each communicator includes two resistors $R_L$ and $R_H$ and two series voltage noise sources $V_{LA}(t)$, $V_{HA}(t)$ and $V_{LB}(t)$, $V_{HB}(t)$ representing the thermal noise of the resistors at Alice and Bob, respectively:

$$S_L(f) = 4kTR_L \qquad (1)$$

$$S_H(f) = 4kTR_H \qquad (2)$$

where $S_L(f)$ is the power spectral density of the voltage noise sources $V_{LA}(t)$, $V_{LB}(t)$ and $S_H(f)$ is the power spectral density of the voltage noise sources $V_{HA}(t)$, $V_{HB}(t)$; $k$ is the Boltzmann constant and $T$ is the temperature.

A switch is used to select one of the resistors to be connected to the wire connecting the two communicators, see Fig. 1. At the beginning of each bit exchange, both Alice and Bob connect a resistor ($R_H$ or $R_L$) to the wire. If both, Alice and Bob connect the higher value resistor, the voltage noise level will be high in the wire. If they both connect the low value resistor, the voltage noise will be low. If they connect different value resistors, the noise level will be intermediate and this is invariant if the resistors are swapped. [1,13]. This level can also be identified by the eavesdropper,

Eve, however she cannot determine who has chosen the low value resistor. For this reason, this is the secure state that can be used for key exchange.

Note that in real applications the noise would be too small, therefore artificial noise generators are used to provide large enough signals in a given frequency band. In this case, the noise equivalent temperature is above $10^9$ K [1]. On the other hand generators can enhance the security and offer new schemes with higher practical security in the non-ideal situations [17].

## *Results*

According to the papers about the KLJN communication method the artificial noise generators are only used to emulate high temperatures, so they must generate Johnson-like noise. Therefore the security proof based on physical laws remains valid [1]. Our approach is in some sense opposite to the previous ones, when security has been proven for the given noise properties. Here we determine what the requirements of noise properties for unconditional security are. On the other hand, our analysis is based on statistical methods instead of physical laws of thermodynamics, therefore it can be more easily understandable for computer engineers and software engineers.

Let us assume that the system is operated in the LH situation, when Alice has switched on the lower value resistor and noise, while Bob uses the higher value resistor and noise as shown in Fig. 1.

In this case Eve measures the following voltage $V_E(t)$ and current $I_E(t)$ (flowing from Bob's side towards Alice) in the wire:

$$V_E(t) = \frac{V_{LA}(t) \cdot R_H + V_{HB}(t) \cdot R_L}{R_L + R_H} \tag{3}$$

and

$$I_E(t) = \frac{V_{HB}(t) - V_{LA}(t)}{R_L + R_H} \tag{4}$$

where $V_{LA}(t)$ and $V_{HB}(t)$ are the voltage noise signals at Alice and Bob, respectively. She can have two hypotheses: the correct one and the opposite. She can calculate the statistics of Alice's voltage noise for both cases. Since she knows the resistor values and the used voltage noise statistics, it is clear, that she will know that her assumption is wrong, if she gets invalid values during her calculations. For the correct assumption she must get correct results of course. Let us see what happens in the case of the wrong hypothesis. In this case Eve assumes that the high value resistor has been chosen by Alice. Therefore she calculates Alice's noise voltage $V_A(t)$ as:

$$V_A(t) = V_E(t) - I_E(t) \cdot R_H = \frac{V_{LA}(t) \cdot R_H + V_{HB}(t) \cdot R_L}{R_L + R_H} + \frac{V_{HB}(t) - V_{LA}(t)}{R_L + R_H} R_H \tag{5}$$

$$V_A(t) = \frac{V_{LA}(t) \cdot 2 \cdot R_H + V_{HB}(t) \cdot (R_H - R_L)}{R_L + R_H} \tag{6}$$

$$V_A(t) = V_{LA}(t) \cdot \frac{2 \cdot R_H}{R_L + R_H} + V_{HB}(t) \cdot \frac{R_H - R_L}{R_L + R_H} \tag{7}$$

The variance is given by the sum of variances:

$$\sigma_A^2 = \sigma_L^2 \cdot \left(\frac{2 \cdot R_H}{R_L + R_H}\right)^2 + \sigma_H^2 \cdot \left(\frac{R_H - R_L}{R_L + R_H}\right)^2 \tag{8}$$

where $\sigma_A^2$ is the variance of $V_A(t)$ and $\sigma_L^2$ and $\sigma_H^2$ are the variances of the voltage noise $V_{LA}(t)$ and $V_{HB}(t)$, respectively.

The communication can only be secure if $\sigma_A = \sigma_H$, otherwise Eve will know that Alice connected the low value resistor and voltage generator to the wire. Substituting this into Eq. (8) yields:

$$\frac{\sigma_H^2}{\sigma_L^2}\left(1 - \left(\frac{R_H - R_L}{R_L + R_H}\right)^2\right) = \left(\frac{2 \cdot R_H}{R_L + R_H}\right)^2 \tag{9}$$

$$\frac{\sigma_H^2}{\sigma_L^2} = \frac{(2 \cdot R_H)^2}{(R_L + R_H)^2 - (R_H - R_L)^2} = \frac{4 \cdot R_H^2}{R_L^2 + 2 \cdot R_L \cdot R_H + R_H^2 - (R_H^2 - 2 \cdot R_L \cdot R_H + R_L^2)} \tag{10}$$

$$\frac{\sigma_H^2}{\sigma_L^2} = \frac{R_H}{R_L} \tag{11}$$

or in other form

$$\frac{\sigma_H}{\sigma_L} = \sqrt{\frac{R_H}{R_L}} \tag{12}$$

Therefore the noise amplitude must depend on the resistance as in the case of thermal noise; it must be proportional to the square root of the resistance. Otherwise the communication is certainly unsecure.

In the following we check how the security depends on the probability distribution of the noise. When the eavesdropper makes the correct assumption, she can calculate the noise signal that Alice is using exactly; therefore she gets the correct probability distribution of course. When she makes the wrong assumption then she obtains:

$$V_A(t) = V_L(t) \cdot \frac{2 \cdot R_H}{R_L + R_H} + V_H(t) \cdot \frac{R_H - R_L}{R_L + R_H} \tag{13}$$

The probability density $p_A(x)$ of $V_A(t)$ is given by the convolution of the probability densities of the two independent terms in Eq. (13). If $p(x)$ is the probability density function with unity variance,

$$\alpha = \sigma_L \cdot \frac{2 \cdot R_H}{R_L + R_H} \tag{14}$$

and

$$\beta = \sigma_H \cdot \frac{R_H - R_L}{R_L + R_H} \tag{15}$$

then

$$\sigma_A^2 = \alpha^2 + \beta^2 \tag{16}$$

where $\sigma_A^2$ is the variance of $V_A(t)$, and

$$p_A(x) = \int_{-\infty}^{\infty} \frac{1}{\alpha} p\left(\frac{x'}{\alpha}\right) \cdot \frac{1}{\beta} p\left(\frac{x - x'}{\beta}\right) dx' \tag{17}$$

If Eq. (12) is satisfied, then $\sigma_A = \sigma_H$, that is needed for secure communication. Furthermore $p_A(x)$ measured by Eve must also be identical to the probability density function $p_H(x)$ of the noise voltages $V_{HA}(t)$ and $V_{HB}(t)$, otherwise Eve can detect that her assumption is wrong. Therefore using Eqs. (16) and (17) $p_A(x)$ can be expressed as

$$p_A(x) = p_H(x) = \frac{1}{\sigma_H} p\left(\frac{x}{\sigma_H}\right) = \frac{1}{\sqrt{\alpha^2 + \beta^2}} p\left(\frac{x}{\sqrt{\alpha^2 + \beta^2}}\right) \tag{18}$$

and finally we get

$$\frac{1}{\sqrt{\alpha^2+\beta^2}} p\left(\frac{x}{\sqrt{\alpha^2+\beta^2}}\right) = \int_{-\infty}^{\infty} \frac{1}{\alpha} p\left(\frac{x'}{\alpha}\right) \cdot \frac{1}{\beta} p\left(\frac{x-x'}{\beta}\right) dx'. \qquad (19)$$

## *Discussion*

Eq. (19) is valid for normal distribution only [19], therefore we can conclude that the noise sources $V_{LA}(t)$, $V_{LB}(t)$ and $V_{HA}(t)$, $V_{HB}(t)$ must have normal distribution and the ratio of their amplitude must be equal to the square root of the ratio of the corresponding resistor values. In other words, Johnson-like noise must be used for the secure key exchange in the KLJN system. Note that although several other distributions – for example Cauchy-distribution – satisfy the condition that the convolution in Eq. (19) does not change the type of distribution, however the finite variance required by energetic considerations is only provided by normal distribution.

It is easy to see that for example random numbers with uniform distribution can't be used for secure communication. In this case Eq. (17) gives a trapezoidal probability density function for $p_A(x)$ as shown on Fig. 2, therefore its deviation from $p_H(x)$ can be very easily detected.

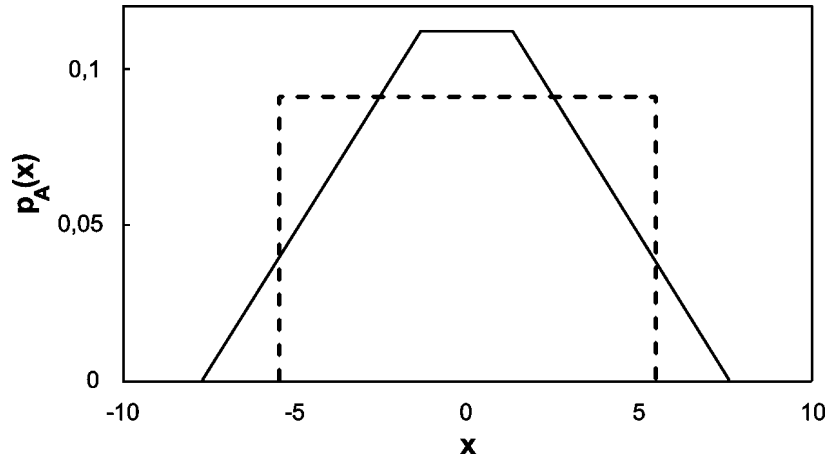

Figure 2. The probability density function $p_A(x)$ in the case of uniform distribution (solid line) strongly differs from $p_H(x)$ (dashed line).

We have developed a simple software application written in LabVIEW that can be used to simulate the KLJN protocol [20]. Normal or uniform distribution can be selected and the values of $R_L$,

$R_H$, amplitude of $V_{LA}(t)$, $V_{LB}(t)$ and $V_{HA}(t)$, $V_{HB}(t)$ can be arbitrarily chosen. The application performs Eve's calculation of $V_A(t)$ for both hypotheses, and plots the corresponding measured amplitudes and probability densities.

## *Limitations and Open Questions*

We have presented a mathematical statistical approach to determine the noise properties and resistor values required for secure communication and the results are in agreement with the original physical approach [1]. On the other hand our work does not address the question of complete security.

Considerable additional work could be carried out to investigate several attack types with similar approach. For example, in practical applications the effect of resistor inaccuracies, wire resistances can also be analyzed using our method; Eq. (8) can be applied to find the difference between the observed and expected variances, $\sigma_A^2$ *and* $\sigma_H^2$, respectively. This means that the information leak due to these inaccuracies can be estimated. On the other hand, if the desired security level is given, the required resistor values and accuracy of the components can be obtained.

Furthermore one can consider correlation properties, bandwidth of the noise sources that is important in practical applications and discussed in several publications.

## *Conclusions*

In this paper we have shown a mathematical statistical approach to find out what kind of noise sources are required for secure communications in the Kirchhoff-Loop-Johnson-Noise unconditionally secure key exchange system. In agreement with the results can be found in the literature we found that the noise amplitude must scale with the square root of the corresponding resistor value and Gaussian noise sources must be used.

Note that our approach can serve as a starting point to quantitatively analyze several attack types in practical applications.

# Acknowledgments

Zoltan Gingl thanks Béla Szentpáli for drawing his attention to the problem. Discussions with Gyula Pap about probability distributions are greatly appreciated.